\documentstyle[11pt,moriond,epsfig]{article}

\def\Journal#1#2#3#4{{#1} {\bf #2}, #3 (#4)}

\def\NIMA{{\em Nucl. Instrum. Methods} A}

\def\be{\begin{equation}}
\def\ee{\end{equation}}
\def\bea{\begin{eqnarray}}
\def\eea{\end{eqnarray}}

\begin{document}
\vspace*{4cm}
\title{CENTRAL DIFFRACTION IN ALICE}

\author{ R. SCHICKER }

\address{Phys. Inst., Philosophenweg 12,\\
69120  Heidelberg, Germany}

\maketitle\abstracts{
The ALICE experiment at the Large Hadron Collider (LHC) at CERN consists 
of a central barrel, a muon spectrometer and of additional detectors 
for trigger and event classification purposes. The low transverse 
momentum threshold of the central barrel gives ALICE a unique opportunity to 
study the low mass sector of central production at the LHC.
I will report on first analysis results of meson production in
double gap events in minimum-bias proton-proton collisions at 
$\sqrt{s}$ = 7 TeV, and will describe a dedicated double gap trigger
for future data taking.}

\section{Introduction}

The ALICE experiment consists of a central barrel and of a forward muon 
spectrometer \cite{Alice1}. 
Additional detectors for trigger purposes and for event classification exist 
outside of the central barrel. Such a geometry allows the investigation 
of many properties of diffractive reactions at hadron colliders, 
for example the measurement of single and double diffractive dissociation 
cross sections and the study of central diffraction. The ALICE physics 
program foresees data taking in pp and PbPb collisions at nominal 
luminosities \mbox{$\cal{L}$ = $5\times 10^{30}cm^{-2}s^{-1}$} and 
\mbox{$\cal{L}$ = $10^{27}cm^{-2}s^{-1}$,} respectively. 
An asymmetric system, pPb, will be measured in the fall of 2012.

\section{The ALICE Experiment}

In the ALICE central barrel, momentum reconstruction and particle 
identification are achieved in the pseudorapidity range $-1.4 < \eta < 1.4$ 
combining the information from the Inner Tracking System (ITS) and the Time 
Projection Chamber (TPC). 

\begin{figure}[htb]                                                            
\begin{center}
\includegraphics[width=0.6\textwidth]{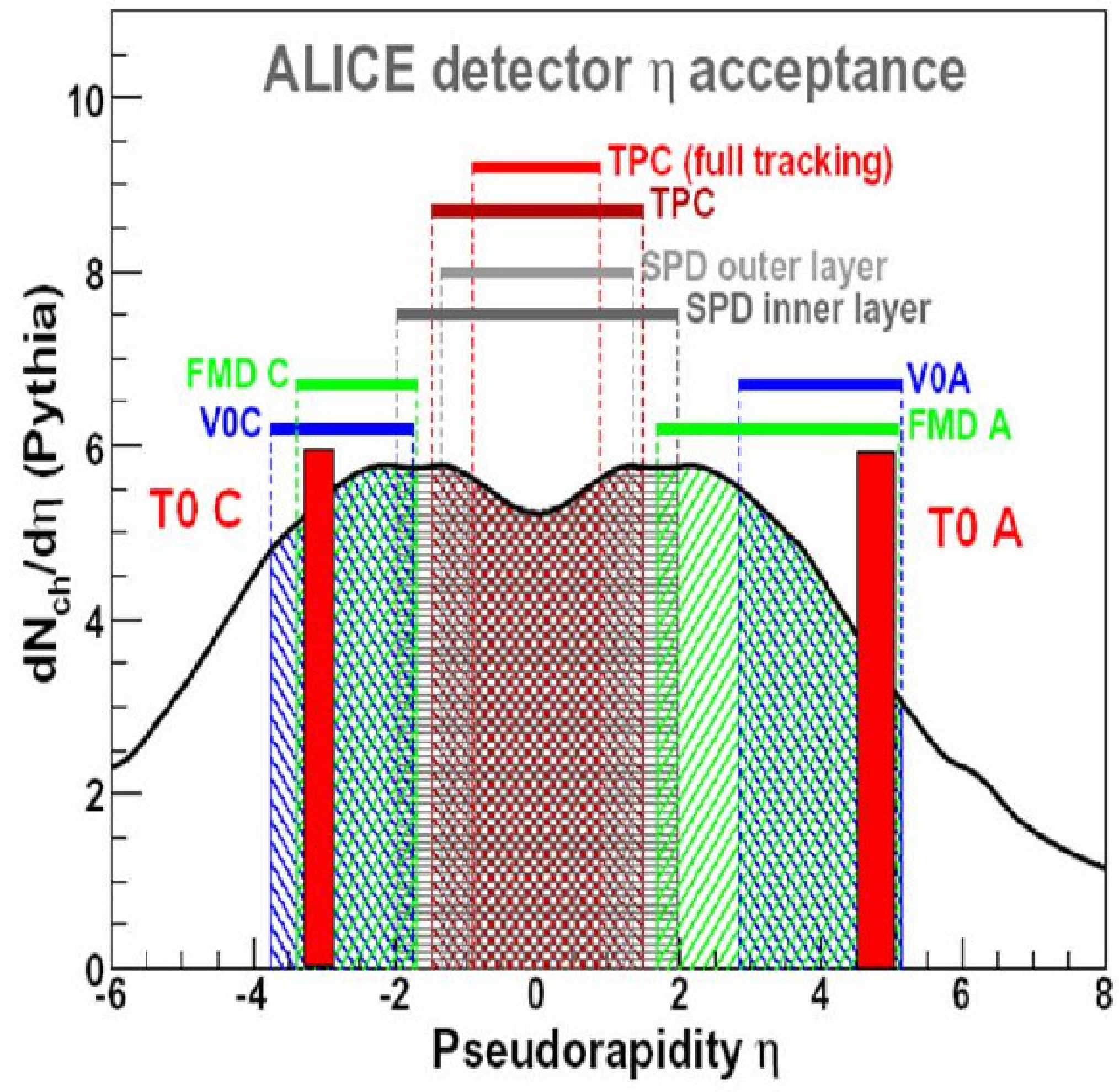}
\caption{Pseudorapidity coverage of the ALICE detectors.}                     
\label{fig:acc}
\end{center}
\end{figure}

In the pseudorapidity range $ -0.9 < \eta <  0.9 $, the information from the 
Transition Radiation Detector (TRD) and the Time of Flight (TOF) system is
also available.
A muon spectrometer covers the range $-4.0<\eta<-2.5$.
At very forward angles, the energy flow is measured by Zero Degree 
Calorimeters (ZDC) \cite{ZDC}.
Detectors for event classification and trigger purposes are located on both 
sides of the ALICE central barrel. First, the scintillator arrays V0A and V0C 
cover the pseudorapidity range $2.8 < \eta < 5.1$ and $-3.7 < \eta < -1.7$, 
respectively. The four- and eightfold  segmentation in pseudorapidity and 
azimuth result in 32 individual counters in each array.
Second, a Forward Multiplicity Detector (FMD) based on silicon strip 
technology covers the pseudorapidity range $1.7 < \eta < 5.1$ and 
$-3.4 < \eta < -1.7$, respectively.
Third, two arrays of Cherenkov radiators T0A and T0C determine 
the time of collisions. Figure \ref{fig:acc} shows the 
pseudorapidity coverage of these detector systems. 
    
\section{Central diffraction in ALICE}        

Central diffractive events are experimentally defined by activity in the       
central barrel and by no activity outside the central  barrel. This 
condition can be implemented at trigger level zero (L0) by defining barrel 
activity as hits in the ITS pixel detector or the TOF system. The gap 
condition is realized by the absence of V0 signals, hence a gap of two 
units in pseudorapidity on either barrel side can be defined at L0. In the 
offline analysis, the information from the V0, T0, FMD, SPD  and TPC 
detectors define the gaps spanning the range $0.9 < \eta < 5.1$ and  
$-3.7 < \eta < -0.9$. Events with and without detector signals in these 
two ranges are defined to be no-gap and  double gap events, respectively.
A rapidity gap can be due either to Pomeron, Reggeon or photon exchange.
A double gap signature can therefore be induced by a combination of these 
exchanges. Pomeron-Pomeron events result in centrally produced states
with quantum numbers C = +1 (C = C-parity) and I = 0 (I = isospin).
The corresponding quantum numbers in photon-Pomeron induced events
are C = -1 and I = 0 or I = 1 \cite{Nachtmann}.

\section{Central meson production in pp-collisions}

In the years 2010-2011, ALICE recorded zero bias and mimimum bias data in 
pp-collisions at a center-of-mass energy of $\sqrt{s}$ = 7 TeV. 
The zero bias trigger was defined by beam bunches crossing at the ALICE
interaction point, while the minimum bias trigger was derived by 
minimum activity in either the ITS pixel or the V0 detector.
Events with double gap topology as described above are contained in this 
minimum bias trigger, hence central diffractive events were analyzed 
from the minimum bias data sample. 

For the results presented below, $3.6 \times 10^8$ minimum bias
events were analyzed.  First, the fraction of events satisfying the 
gap condition described above was calculated.  This fraction was 
found to be about $2 \times 10^{-4}$.  Only runs where  this fraction 
was calculated to be within 3 $\sigma$ of the average value of the 
corresponding distribution were further analyzed.  
This procedure resulted in about $7 \times 10^{4}$ double gap events.
As a next step, the track multiplicity in the pseudorapidity 
range $ -0.9 < \eta < 0.9$  was evaluated .

\begin{figure}[htb]
\begin{center}
\includegraphics[width=0.7\textwidth]{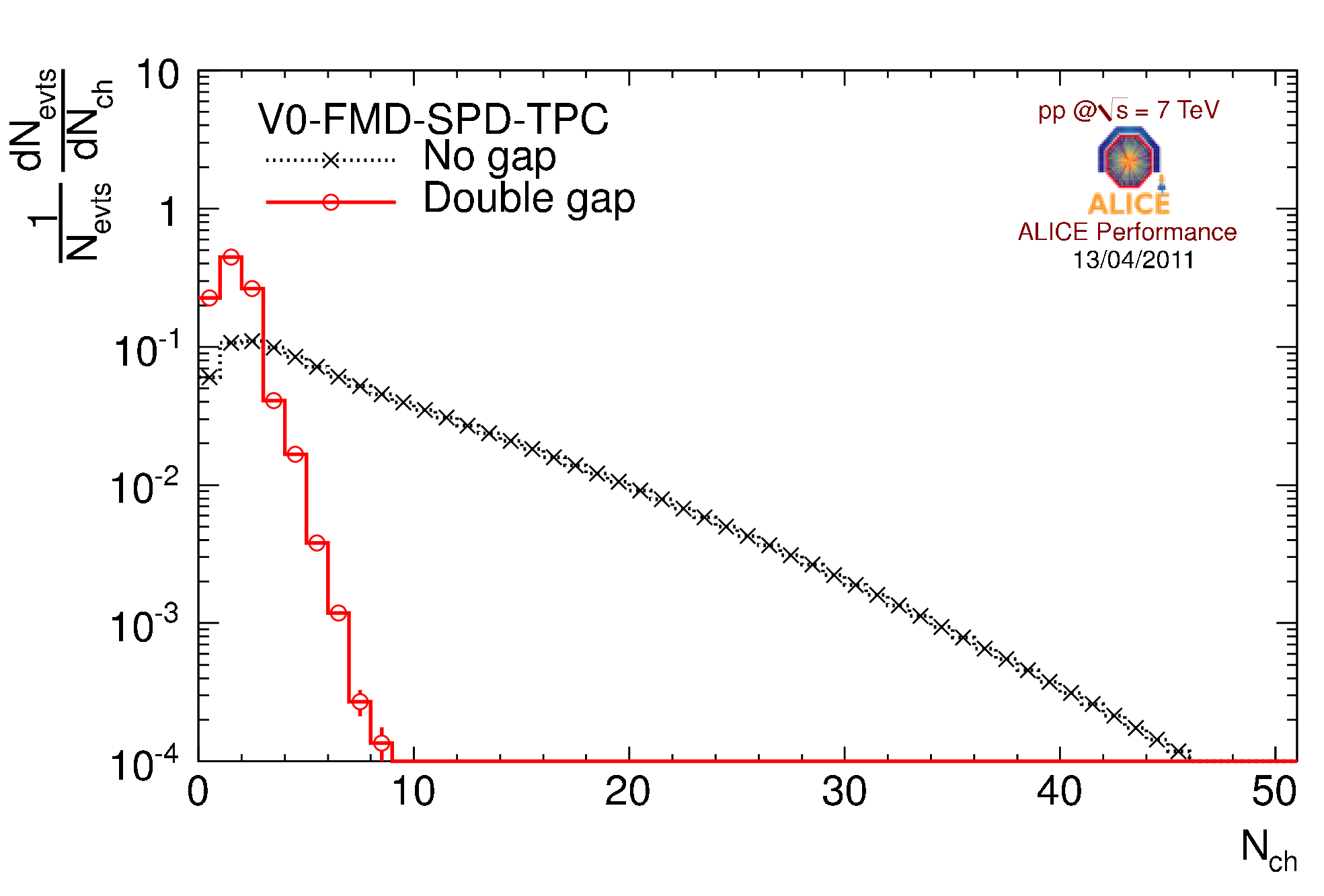}
\caption{Track multiplicity within the pseudorapidity range 
-0.9 $ < \eta < $ 0.9 for no-gap and double gap events.}
\label{fig:dgap_mult}
\end{center}
\end{figure}

Figure \ref{fig:dgap_mult} shows the track multiplicity in the pseudorapidity 
range \mbox{$ -0.9 < \eta < 0.9 $}  for double and 
no-gap events. Very low transverse momentum
tracks never reach  the TPC which results in events with track multiplicity 
zero. The multiplicity distributions of the double and no-gap events clearly 
show different behaviors as demonstrated  in Figure \ref{fig:dgap_mult}. 

\begin{figure}[htb]
\begin{center}
\includegraphics[width=0.7\textwidth]{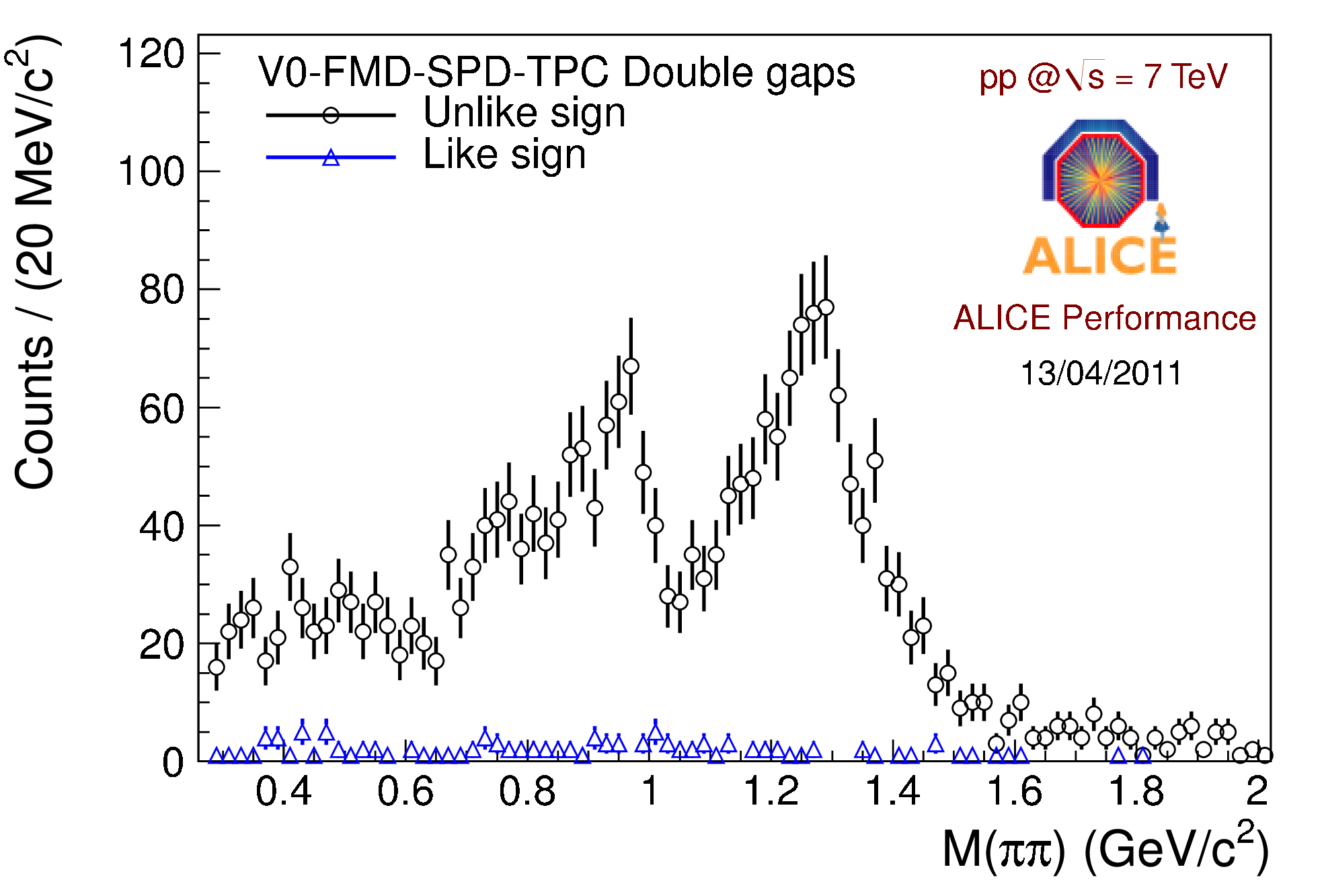}
\caption{Invariant mass distribution of like and unlike sign pion pairs.}
\label{fig:dgap_mass_lu}
\end{center}
\end{figure}

The specific energy loss dE/dx as measured by the 
TPC in combination with the TOF detector information 
identifies pions with transverse momenta p$_{T} \geq$ 300 MeV/c. 
The events with exactly two pions are selected, and the invariant mass
of the pion pairs is shown in \mbox{Figure \ref{fig:dgap_mass_lu}.} 
These pion pairs can be of like or unlike sign charge. Like sign pion 
pairs can arise from two pion pair production with loss of one pion 
of same charge in each pair, either due to the low p$_T$ cutoff described above,
or due to the finite pseudorapidity coverage of the detectors used for 
defining the rapidity gap. For charge symmetric detector
acceptances, the unlike sign pairs contain the signal plus background,
whereas the like sign pairs represent the background.
From the two distributions shown in Figure \ref{fig:dgap_mass_lu},
the background is estimated to be less than 5\%.

\begin{figure}[htb]
\begin{center}
\includegraphics[width=0.7\textwidth]{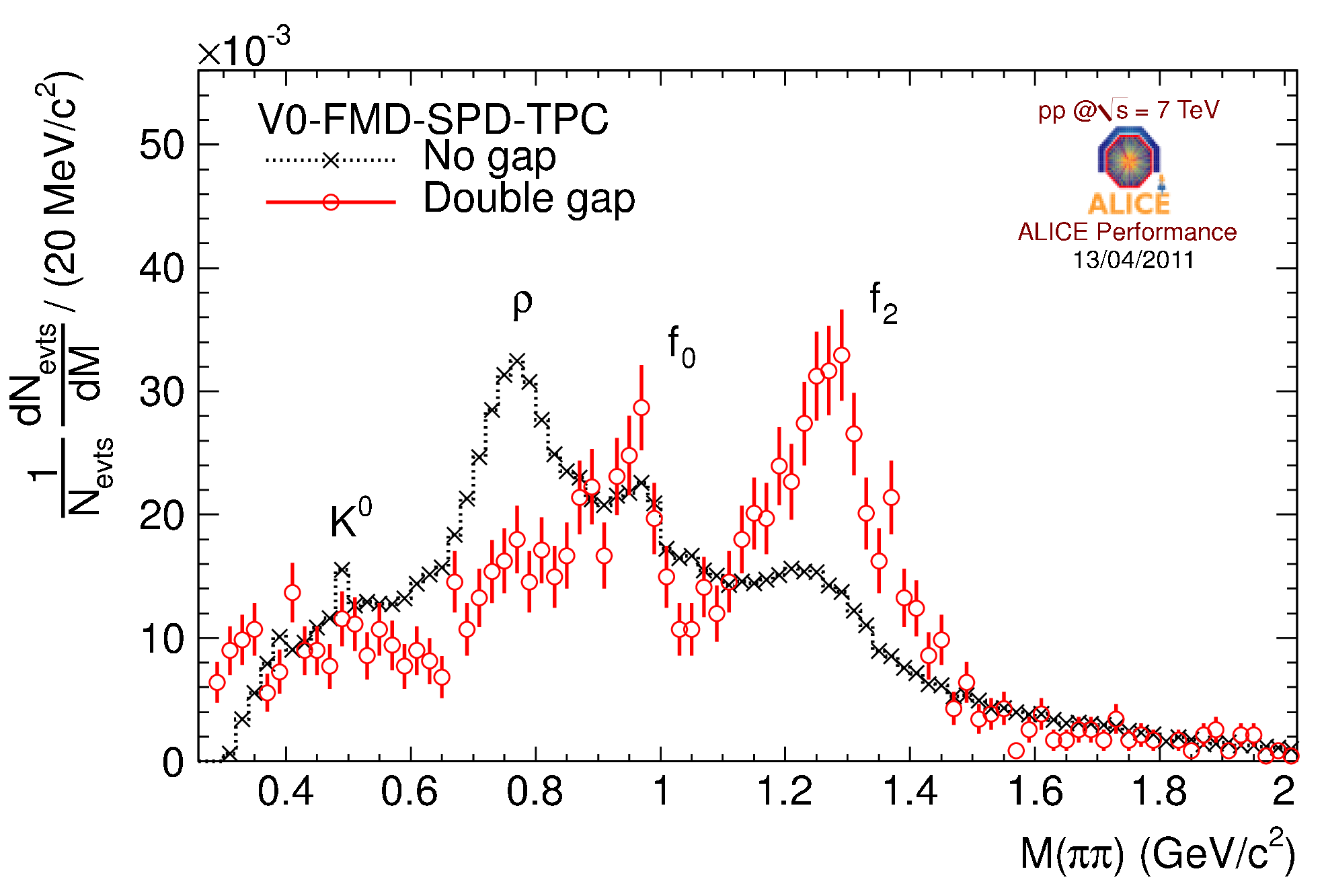}
\caption{Pion pair invariant mass distribution for double and 
for no-gap events.}
\label{fig:dgap_mass}
\end{center}
\end{figure}

Figure \ref{fig:dgap_mass} displays the normalized background corrected
pion pair mass for double and no-gap events. 
The particle identification by the TOF detector
requires the single track transverse momentum p$_T$ to be larger than 
about 300 MeV/c. This single track p$_{T}$ cut introduces a significant
acceptance reduction for pair masses $M(\pi\pi) \leq$ 0.8 $GeV/c^2$
at low pair p$_{T}$. The distributions shown are not acceptance corrected. 
In the no-gap events, structures are seen from $K^{0}_{s}$ and 
$\rho^{0}$-decays. Two additional structures are associated with f$_0$(980) 
and f$_2$(1270) decays. In the double gap distribution, the $K^{0}_{s}$ 
and $\rho^{0}$ are highly suppressed while the f$_0$(980) and f$_2$(1270) 
with quantum numbers J$^{PC}$ = (0, 2)$^{++}$ are much enhanced. This 
enhancement of \mbox{J$^{PC}$ = J$^{++}$} states is evidence that the 
double gap condition used for analysing the minimum bias data sample 
selects events dominated by double Pomeron exchange.

\section*{Acknowledgments}
This work is supported in part by German BMBF under project 06HD197D and 
by WP8 of the hadron physics program of the 7$^{th}$  EU program period.

\end{document}